\documentclass{article}
\pdfoutput=1
\usepackage{spconf,amsmath,graphicx}
\usepackage{amsmath,amsthm,amssymb,latexsym,color, dsfont}
\usepackage{wrapfig}
\usepackage[linesnumbered,ruled]{algorithm2e}
\usepackage{caption}
\usepackage{subfig}
\usepackage{booktabs}
\usepackage{hyperref}

\DeclareMathOperator*{\argmin}{arg\,min}

\usepackage{mathabx}

\usepackage{bbm}
\usepackage{mathtools}
\usepackage{comment}

\renewcommand{\epsilon}{\varepsilon}

\newtheorem{remark}{Remark}

\DeclareMathOperator*{\Tr}{Tr}

\usepackage{xcolor}

\title{On audio enhancement via online non-negative matrix factorization}
  \name{Andrew Sack$^{\star}$, \quad Wenzhao Jiang$^{\dagger}$ \quad Michael Perlmutter$^{\star}$ \quad Palina Salanevich$^{\ddagger}$ \quad Deanna Needell$^{\star}$\thanks{The authors were partially supported by NSF DMS $\#2011140$ and NSF DMS $\#1909457$.}}
\address{$^{\star}$ Univ. California, Los Angeles, Department of Mathematics, Los Angeles, CA, USA \\
$^{\dagger}$Univ. Science and Technology of China, Department of Mathematics, Hefei, Anhui, China\\
$^{\ddagger}$Utrecht University, Mathematical Institute, the Netherlands}

\begin{document}
%

\maketitle
%
\begin{abstract}
We propose a method for noise reduction, the task of producing a clean audio signal from a recording corrupted by additive noise. Many common approaches to this problem are based upon applying non-negative matrix factorization to spectrogram measurements. These methods use a noiseless recording, which is believed to be similar in structure to the signal of interest, and a pure-noise recording to learn  dictionaries for the true signal and the noise.  
One may then construct an approximation of the true signal by projecting the corrupted recording on to the clean  dictionary. 
In this work, we build upon these methods by proposing the use of  \emph{online} non-negative matrix factorization for this problem. This method is more memory efficient than traditional non-negative matrix factorization and also has potential applications to real-time denoising. 
\end{abstract}

\section{Introduction}
\label{sec:intro}

Audio is frequently recorded in a noisy environment, such as the outdoors or a stadium filled with screaming fans.  This motivates the task of \textit{noise reduction}, i.e.,  removing noise from a corrupted transmission and returning a clean recording.  This has a myriad of applications, such as enhancing the quality of a concert recording or making a speech easier to parse for automated transcription. Therefore, a number of approaches to this problem have been studied, including Wiener filters \cite{surhone2010wiener},  spectral noise gates \cite{Kiapuchinski2012}, and deep neural networks \cite{Bai2018}. 

One approach that has seen a lot of use in recent years is dictionary learning via \emph{non-negative matrix factorization} (NMF)~\cite{Schmidt2007, Wilson2008, shimada2019unsupervised, schmidt2006single, schulze2021sparse}.
In the NMF method, one aims to factorize a non-negative \emph{data matrix} $\mathbf{X}\in \mathbb{R}_{\ge 0}^{d\times n}$  as a product of a \emph{dictionary matrix} $\mathbf{W}\in \mathbb{R}_{\ge 0}^{d\times k}$ and a \emph{code matrix} $\mathbf{H}\in \mathbb{R}_{\ge 0}^{k\times n}$,  with $k\ll d$. We typically interpret columns 
of the data matrix $\mathbf{X}$ as the data points, and the columns 
of the dictionary matrix $\mathbf{W}$ as the atoms of a dictionary, such that every data point 
can be represented as a non-negative linear combination of dictionary atoms. Furthermore, one often desires that each data point is represented using only a  few dictionary atoms, which can be mathematically formalized as a sparsity requirement on the code matrix~$\mathbf{H}$.

A major advantage of NMF is its interpretability. For example, when applied to pictures of human faces, the atoms of the dictionary $\mathbf{W}$ produced by NMF will resemble, e.g, a person's eyes or ears~\cite{lee1999learning}. This is in contrast to other factorization approaches, such as principle component analysis, which yield hard to interpret ``eigenfaces''~\cite{navarrete2002analysis}. This  qualitative difference is explained by the non-negativity constraint. Since there are no negative terms, cancellation cannot occur, and thus, dictionary atoms are forced to concentrate around salient features. {In the context of audio denoising, the dictionary obtained by NMF factorization is expected to pick up structural features of the clean signal 
so that the signal and the noise components of the noisy recording will be represented by different dictionary atoms.}

In most cases, an exact factorization $\mathbf{X}=\mathbf{W}\mathbf{H}$ is not possible.  Instead, one aims to obtain a close  approximation of the data matrix $\mathbf{X}$ by a product $\mathbf{WH}$ by minimizing a loss function $\mathcal{L}(\mathbf{X},\mathbf{W}, \mathbf{H})$ subject to the constraint that $\mathbf{W}$ and $\mathbf{H}$ have non-negative entries. In this paper, we will measure the approximation error via the Frobenius norm and use $\ell^1$ regularization  to promote sparsity of the code matrix. {With these choices, we are able to obtain our code and dictionary matrices via an explicit multiplicative update formula.} 

In this paper, we obtain the data matrix $\mathbf{X}$ by taking the \emph{spectrogram} of the input signal $\mathbf{x}(t)$, i.e., the magnitude of its short-time Fourier transform. Such measurements are commonly used in audio processing tasks (see, e.g., \cite{dieleman2014end}) and encode information about which frequencies are active at each point of time in a recording. 
Our method is motivated by the observation that, unlike noise, most natural sounds, such as voices and musical instruments, have only a few dominant frequencies active at each point in time. Thus, one may hope to obtain a dictionary in which ``signal atoms'' can be readily distinguished from the ``noise atoms''.  In particular, given prior information on what the signal and the noise are like, one can train two dictionaries $\mathbf{W_{\text{S}}}$ and $\mathbf{W_{\text{N}}}$.  Then, if $\mathbf{X}$ and $\mathbf{S}$ denote the spectrograms of the noisy signal and the clean signal, respectively, one can expect that  $\mathbf{W_{\text{S}}}$ can sparsely code the spectrogram of the clean signal $\mathbf{S}$ but not $\mathbf{N}\coloneqq \mathbf{X}-\mathbf{S}$.

In this paper, we use Online NMF (ONMF), an algorithm developed for streaming data or for situations where the data set is too large to store in local memory. In the latter case, ONMF alleviates the  memory burden by allowing one to load only a portion of the data set at a time.  As with traditional NMF, we learn a dictionary matrix $\mathbf{W}$ where the columns of $\mathbf{X}$ can be well-approximated as non-negative linear combinations of the columns of $\mathbf{W}$.  However, in this context, we view the columns of $\mathbf{X}$ as samples of some probability distribution and we view the dictionary $\mathbf{W}$ as learning the essential components of this distribution. As discussed in \cite{lyu2020applications}, there is a natural application of ONMF to non-negative time-series data where the columns are interpreted as the terms of a vector-valued stochastic process.   

\subsection{Contribution}\label{sec:contribution}
This paper builds upon previous work  applying NMF to spectrogram measurements for the purpose of audio separation (see, e.g. \cite{fevotte2018single, Wilson2008}). Instead of the traditional NMF approach developed in these previous works, we use \textit{online} non-negative matrix factorization. This is motivated by several considerations. 

{First, when using ONMF, one views the spectrogram as a time-series of vectors, where each vector indicates the active frequencies at a specific time. By sampling batches from this time-series, ONMF learns a dictionary that is better suited to represent phonemes and musical chords, compared to the dictionary obtained using NMF, which does not exploit the time-frequency interpretation of the spectrogram.
Therefore, one might hope that ONMF will learn different atoms from traditional NMF and achieve better reconstruction.} 
Indeed, based on our numerical experiments in Section~\ref{sec:exp}, the dictionary learned by ONMF is different than the one learned by traditional NMF and our ONMF-based denoising algorithm exhibits superior performance. 

Secondly, ONMF does not require one to store the entire spectrogram $\mathbf{X}$, but instead works with smaller matrices obtained by subsampling the columns of $\mathbf{X}$. Therefore, ONMF is significantly more memory-efficient than traditional NMF. 

Lastly, unlike traditional NMF which requires the entire recording to be known in advance, ONMF has potential applications to the real-time denoising of streaming audio.  As a simple motivating example, consider the streaming broadcast of a concert.  A microphone near the band will observe a mixed signal of the band and audience.  We can learn a dictionary for the clean signal from the band's studio recordings, and a microphone placed in the audience will observe a signal with very heavy noise. Using ONMF, one could use the sound picked up by the audience microphone to a learn dictionary for the noise in real time and use this dictionary to denoise the recording picked up by the stage microphone. 

\medskip

The rest of this paper is organized as follows. In Section~\ref{sec:method}, we explain how to apply ONMF to the noise reduction problem. In Section~\ref{sec:exp}, we present experimental results demonstrating the utility of the ONMF approach and its advantages over traditional NMF. Lastly, in Section~\ref{sec:conc}, we provide a brief conclusion and discuss future research directions.

\section{Method} \label{sec:method}

We assume that the observed signal $\mathbf{x}(t)$ can be written as $$\mathbf x(t) = \mathbf s(t) +\mathbf n(t)$$ where $\mathbf s$ is the clean signal and  $\mathbf n$ is noise.  Let $\mathbf{X}$ and $\mathbf{S}$ denote the spectrograms of $\mathbf x(t)$ and $\mathbf s(t)$, respectively, and denote~$\mathbf{N} = \mathbf X - \mathbf S$. The spectrogram is not linear, but heuristically, we will think of $\mathbf{N}$ as the spectrogram of the noise.

We further assume that we have priors for signals $\mathbf s$ and $\mathbf n$.  More precisely, if, e.g., $\mathbf x(t)$ is a noisy recording of a person talking, then we assume to have access to a clean sample $\mathbf{s}'(t)$ of that person talking and also a sample of the noise $\mathbf{n}'(t)$ recorded in the same environment or sampled from the same probability distribution as $\mathbf{n}(t)$.  Let $\mathbf S'$ and $\mathbf N'$ denote the spectrograms of $\mathbf{s}'(t)$ and $\mathbf{n}'(t)$, respectively. We take $\mathbf{W}_S$ and $\mathbf W_N$ to be dictionaries learned using NMF approach from $\mathbf S'$ and $\mathbf N'$ by minimizing the loss function \eqref{eqn: objective}. 

\begin{algorithm}[h]
\SetKwInOut{Input}{Input}
\SetKwInOut{Output}{Output}

\Input{Data matrix $\mathbf X$; initialization $\mathbf{W}_0,\mathbf{H}_0$}
\Output{$\mathbf{W}_K,\mathbf{H}_K$, s.t. $\mathbf X  \approx \mathbf{W}_0\mathbf{H}_0$}
\smallskip
\For{$k=1, \dots, K$}{
Update Code Matrix: $\displaystyle (\mathbf{H}_k)_{i,j}=(\mathbf{H}_{k-1})_{i,j}\frac{(\mathbf{W}_{k-1}^T\mathbf{X})_{i,j}}{(\mathbf{W}_{k-1}^T\mathbf{W_{k-1}}\mathbf{H}_{k-1})_{i,j}} $\\
\medskip

Update Dictionary Matrix:
$\displaystyle(\mathbf{W}_k)_{i,j}=(\mathbf{W}_{k-1})_{i,j}\frac{(\mathbf{X}\mathbf{H}_{k}^T)_{i,j}}{(\mathbf{W}_{k-1}\mathbf{H}_{k}\mathbf{H}_{k}^T)_{i,j}}$
}
\caption{NMF via Multiplicative Updates}\label{alg: NMF}
\end{algorithm}

As the columns of $\mathbf W_S$ are trained to represent the pure signal, we expect that they can sparsely code $\mathbf S$ but not $\mathbf N$. 
In order to decompose $\mathbf X$ into $\mathbf S + \mathbf N$, we concatenate $\mathbf W_S$ and $\mathbf W_N$ to form $\mathbf W\coloneqq (\mathbf{W}_S, \mathbf{W}_N)$ and 
  obtain a coding matrix $\mathbf{H}\coloneqq (\mathbf{H}_S^T,\mathbf{H}_N^T)^T$ by minimizing the loss function 
\begin{align}\label{eqn: objective}
\mathcal{L}(\mathbf{X},\mathbf{W},\mathbf{H})\coloneqq    \frac{1}{2}\mathbf{\|\mathbf{X}-\mathbf{W}\mathbf{H}\|^2_F + \alpha\|\mathbf{H}\|_1}, 
\end{align}
for a suitably chosen regularization parameter $\alpha>0$.  This problem can be solved in a variety of ways including, e.g., the multipicative update scheme described in Algorithm \ref{alg: NMF} (see, e.g., \cite{fevotte2018single}).
After solving this minimization problem, we will have 
$\mathbf S \approx \mathbf W_S \mathbf H_S $ and~${\mathbf N \approx \mathbf W_N \mathbf H_N}$.  

One of the drawbacks of this NMF-based noise reduction approach is that  minimization of $\mathcal{L}(\mathbf S', \mathbf W_S, \mathbf H_S')$ and $\mathcal{L}(\mathbf N', \mathbf W_N, \mathbf H_N')$ requires the entire matrices $\mathbf S'$ and $\mathbf N'$ to be loaded in memory and to be available ahead of time.  ONMF, on the other hand, is much more memory efficient, as it does not require loading the entire data matrix at the same time. Furthermore, it does not require that $\mathbf{S}'$ and $\mathbf{N}'$ are known in advance, which makes it possible to use in the online settings. 

Our method is based on sampling columns of $\mathbf{X}$, which we interpret as ``time slices'' of the spectrogram, and organizing them into ''time sample'' submatrices. We then iteratively update the dictionary matrix to obtain a close approximation of all the sampled submatrices. 
More precisely, let $T$ be some large integer and choose $m\ll n$. For $1\leq t \leq T$, we let $\mathbf{X}_t\in\mathbb{R}^{d\times m}$ be an $d\times m$ matrix obtained by randomly selecting $m$ columns of $\mathbf{X}$. In ONMF, one iteratively learns the best factorization $\mathbf{X}_t\approx \mathbf{W}_t \mathbf{H}_t$. On each iteration, one first finds $\mathbf{H}_t$ which minimizes $\mathcal{L}(\mathbf{X}_t,\mathbf{W}_{t-1},\mathbf{H})$ and then finds $\mathbf{W}_t$ which minimizes the average value of $\mathcal{L}(\mathbf{X}_s,\mathbf{W},\mathbf{H}_{s})$, $1\leq s \leq t$. Unfortunately, the straightforward update rule of
\begin{equation}\label{eqn: intuiutive}
    \mathbf{W}_t=\argmin_{\mathbf{W}\geq 0} \frac{1}{t}\sum_{s=1}^t \mathcal{L}(\mathbf{X}_s,\mathbf{W},\mathbf{H}_s)
\end{equation}
requires storing all of the matrices $\mathbf{X}_s, 1\leq s \leq t$. This creates a  considerable memory burden. In \cite{mairal2010online}, the authors were able to alleviate it by aggregating all of the relevant past information from the first $t-1$ steps into two aggregation matrices $\mathbf{A}_t$ and $\mathbf{B}_t$. As detailed in Algorithm \ref{alg: ONMF}, this allows one to compute $\mathbf{W}_t$ via multiplicative updates without storing all of the $\mathbf{X}_s$, $1\leq s\leq t-1.$ Moreover, the authors show that this more complicated update rule is equivalent to the intuitive update rule \eqref{eqn: intuiutive} and also provide theoretical convergence guarantees for i.i.d.\ data. These convergence guarantees were subsequently extended to Markovian data in \cite{lyu2020online}. 

\begin{algorithm}[h]
\SetKwInOut{Input}{Input}
\SetKwInOut{Output}{Output}

\Input{data matrix $\mathbf X$; initialization $\mathbf{W}_0$, $\mathbf{A}_0,\mathbf{B}_0=\mathbf{0}$, regularized parameter $\alpha >0$}
\Output{$\mathbf{W}_T,\mathbf{H}_T$, s.t. $\mathbf X  \approx \mathbf{W}_0\mathbf{H}_0$}
\smallskip
\For{$t=1, \dots, T$}{
Update Sparse Code Matrix: $\displaystyle \mathbf{H}_t=\argmin_{\mathbf{H}\geq 0} \|\mathbf{X}_t-\mathbf{W}_{t-1}\mathbf{H}\|_F^2+\alpha\|\mathbf{H}\|_1$\\

\medskip

Aggregate Past Information:
$\displaystyle \mathbf{A}_t=\frac{1}{t}\left((t-1)\mathbf{A}_{t-1}+\mathbf{H}_t\mathbf{H}_t^T\right)$\newline

$~~\displaystyle \mathbf{B}_t= \frac{1}{t}\left((t-1)\mathbf{B}_{t-1}+\mathbf{H}_t\mathbf{X}_t^T\right)$\\

\medskip

Update Dictionary Matrix:\newline
$\displaystyle \mathbf{W}_t=\argmin_{\mathbf{W}\geq 0}\frac{1}{2}\Tr(\mathbf{W} \mathbf{A}_t\mathbf{W}_t^T) - \Tr(\mathbf{B}_t\mathbf{W})$
}
\caption{ONMF}\label{alg: ONMF}
\end{algorithm}

\begin{remark}
Since matrices $\mathbf{X}_t$ are obtained by randomly sampling the columns of $\mathbf X$, we naturally obtain a sequence of i.i.d.\ data matrices as analyzed in \cite{mairal2010online}. However, in the setup of online audio processing, it is reasonable  to model the columns of $\mathbf{S}'$ as a vector-valued Markov chain. In this case, if one also models the columns of $\mathbf{N}'$ as being either i.i.d. or  Markovian, 
then one may view the columns $\mathbf{X}$ as Markovian. Therefore, in light of the theoretical guarantees produced in \cite{lyu2020applications}, one can modify our proposed algorithm by selecting $\mathbf{X}_t$ to be $m$ \textit{consecutive} columns of $\mathbf{X}$ rather than $m$ columns chosen uniformly at random. Doing so would eliminate the requirement of having access to the entire matrix $\mathbf{S}'$ and would allow for the potential real-time denoising of streaming audio, as discussed in Section \ref{sec:contribution}.
\end{remark}

By applying the ONMF algorithm described in Algorithm~\ref{alg: ONMF}  to $\mathbf S'$ and $\mathbf N'$, we obtain dictionaries $\mathbf W_S$ and $\mathbf W_N$, which we concatenate into $\mathbf{W}\coloneqq (\mathbf{W}_S,\mathbf{W}_N)$. Then we find a sparse coding matrix $\mathbf{H}=(\mathbf{H}_S^T,\mathbf{H}_N^T)^T$ such that $\mathbf X\approx \mathbf{W}\mathbf{H}$ and define our estimates of the $\mathbf{S}$ and $\mathbf{N}$ by 
\begin{equation*}
    \mathbf{S}_\text{est}\coloneqq \mathbf{W}_S\mathbf{H}_S\quad\text{and}\quad\mathbf{N}_\text{est}\coloneqq \mathbf{W}_N\mathbf{H}_N
\end{equation*}
Next, we apply a post-processing step to enforce that \linebreak${\mathbf X = \widetilde{\mathbf{S}}_\text{est} + \widetilde{\mathbf{N}}_{\text{est}}}$.    
In particular, we set 
\begin{align*}
    (\widetilde{\mathbf{S}}_\text{est})_{i,j}\coloneqq &
    \frac{(\mathbf{S}_\text{est})_{i,j}(\mathbf{X})_{i,j}}{(\mathbf{S}_{\text{est}})_{i,j}+(\mathbf{N}_{\text{est}})_{i,j}},
\end{align*}
and define $\widetilde{\mathbf{N}}_\text{est}$ similarly. 
As the experiments show, this step greatly enhances the quality of our recovered audio. The matrix  $\widetilde{\mathbf{S}}_\text{est}$ is our estimate of \textit{magnitudes} of the STFT of $\mathbf{s}(t)$. We estimate the \textit{phases}  by setting them equal to the phases of the STFT of $\mathbf{x}(t)$. Finally, we obtain the estimate apply an inverse STFT to get an estimate of $\mathbf{s}(t).$

\section{Experiments} \label{sec:exp}

In our experiments, we will apply both OMNF and traditional NMF to the noise reduction problem and show that ONMF exhibits superior performance.\footnote{Our code is available at \url{https://github.com/Jerry-jwz/Audio-Enhancement-via-ONMF}.} We consider signals corrupted by both synthetically produced Gaussian white noise and signals corrupted by real-world noise produced by a Levoit-H132 air purifier.
We evaluate a denoising method performance using three standard accuracy measures: the signal-to-distortion ratio (SDR), the signal-to-interference ratio (SIR) and the signal-to-artifacts ratio (SAR). We refer the reader to \cite{vincent2006measures} for the definitions. 

 In both methods, we let the signal dictionary $\mathbf{W}_S$ have 50 columns and let the noise dictionary $\mathbf{W}_N$ have  10. In the training stage, we construct our initial dictionary by setting each entry to have i.i.d.\ random entries uniformly distributed between 0 and 1. After each iteration, we renormalize our columns to have unit $l^2$-norm. In the sparse coding stage, we set the regularizer parameter $\alpha$ equal to 100. 
 
 In Algorithm \ref{alg: NMF}, we apply the stopping criterion:
 \begin{align*}
 \frac{|\mathcal{L}(\mathbf{X},\mathbf{W}_{k},\mathbf{H}_{k}) - \mathcal{L}(\mathbf{X},\mathbf{W}_{k-1},\mathbf{H}_{k-1})|} {\mathcal{L}(\mathbf{X},\mathbf{W}_0,\mathbf{H}_0)} < 10^{-4}
 \end{align*}
 In Algorithm \ref{alg: ONMF}, we set $T=100$ and for $1\leq t \leq T$ we randomly selecting 100 columns of $\mathbf{X}$ to form $\mathbf{X}_t.$
 In Tables \ref{table: three measures} and \ref{table: three measures real-noise}, we summarize SDR, SIR, and SAR of the original, noisy signal as well as the signals obtained via the NMF and ONMF based denoising approaches. These values of SDR, SIR, and SAR are inversely proportional to the noise norm, so larger values imply are better quality of denoising. 
 
 \begin{table}[h]
\begin{center}
\begin{tabular}{cccc}
  \toprule
  Method & SDR & SIR & SAR \\
  \midrule
  NMF & 19.43 & 31.42 & 19.72 \\
  ONMF & 22.70 & 53.45 & 22.70 \\
  ORIGINAL & 9.75 & 9.76 & 37.41 \\
  \bottomrule
\end{tabular}
\end{center}
\caption{Performance measures with white noise. }\label{table: three measures}
\end{table}

\begin{table}[h]
\begin{center}
\begin{tabular}{cccc}
  \toprule
  Method & SDR & SIR & SAR \\
  \midrule
  NMF & 9.46 & 13.90 & 11.63 \\
  ONMF & 10.41 & 13.11 & 13.95 \\
  ORIGINAL & 5.91 & 5.91 & 286.50 \\
  \bottomrule
\end{tabular}
\end{center}
\caption{Performance measures with real-world noise.}\label{table: three measures real-noise}
\end{table}
 
\begin{figure}[h]
    \captionsetup[subfigure]{labelformat=simple}
    \centering
    \subfloat[Original clean speech]{\includegraphics[width=125pt]{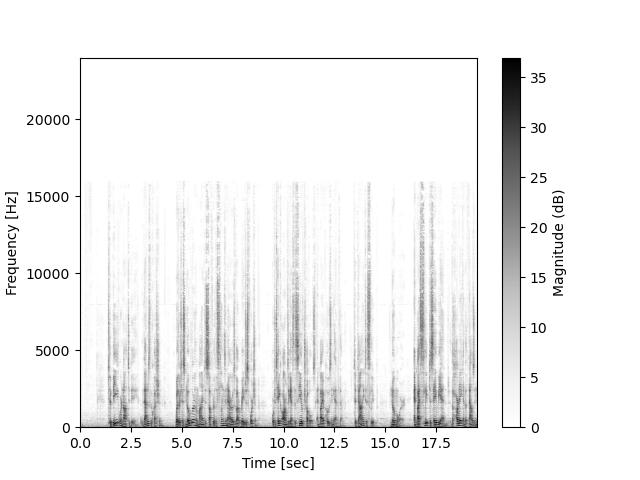}} 
    \subfloat[Original noisy speech]{\includegraphics[width=125pt]{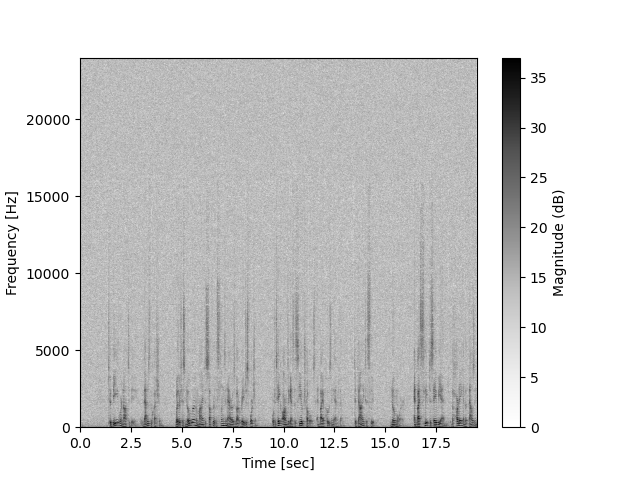}} \\
    \subfloat[NMF-based denoising]{\includegraphics[width=125pt]{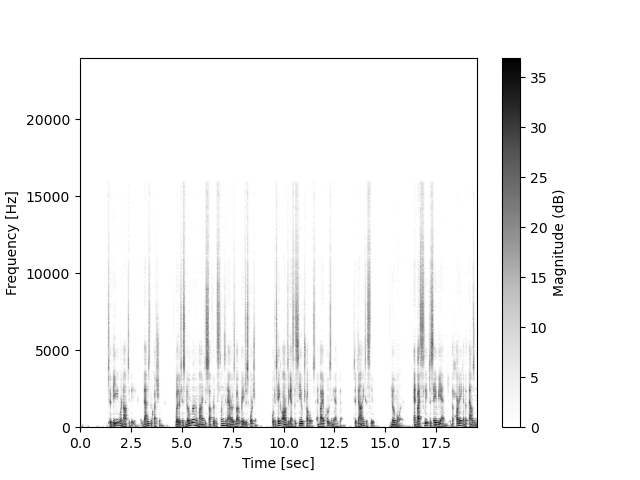}}
    \subfloat[ONMF-based denoising]{\includegraphics[width=125pt]{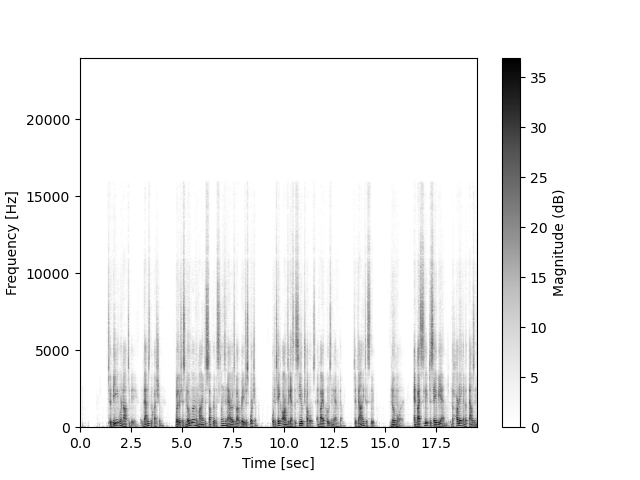}}
    \caption{Spectrograms of the a) original clean signal,
    b) noisy signal corrupted by  white noise c)  signal denoised via traditional NMF d)  signal denoised via ONMF.} 
    \label{fig: three spectrograms}
\end{figure}

Figure \ref{fig: three spectrograms} qualitatively illustrates the performance of NMF- and ONMF-based denoising algorithms. It shows a plot of the original clean and noisy signal spectrograms, as well as those reconstructed by the two algorithms from a signal corrupted by white noise. We observe both qualitatively and quantitatively that our reconstruction algorithm based on ONMF outperforms the one based on traditional NMF.

Between the two experiments, we see that the ONMF-based method outperforms its NMF-based counterparts on five out of six metrics and is only slightly worse with respect to SIR in the case of real-world noise. In both methods, we observe that the algorithm appears to introduce artifacts and the  SAR  decreases during the denoising process (quite significantly in the case of real-world noise). 

We also investigate the role of our regularization parameter $\alpha$. In Table \ref{table: SIRs for different sparsity params}, we plot the SIR for different values of $\alpha$ while using 50 columns in the signal dictionary and 10 columns in the noise dictionary. Similar experiments with differing numbers of columns and using SDR and SAR in place of SIR indicate essentially the same dependencies. 
\begin{table}[h]
\begin{center}
\begin{tabular}{cccccc}
  \toprule
  $\alpha$ & 50 & 60 & 70 & 80 & 90 \\
  \midrule
  SIR & 52.71 & 59.54 & 80.77 & 58.96 & 53.70 \\
  \bottomrule
\end{tabular}
\end{center}
\caption{SIRs for different regularization parameters.}\label{table: SIRs for different sparsity params}
\end{table}
These results suggest that the accuracy of denoising improves as $\alpha$ increases  until a certain threshold value, after which the accuracy deteriorates. This is because we need our code to have enough nonzero terms so that $\mathbf{W}_S$ can be used to approximate $\mathbf{S},$ but not so many that it can be used to approximate $\mathbf{N}$. 
\section{Conclusion} \label{sec:conc}

In this paper, we have proposed a novel method of noise reduction based on online non-negative matrix factorization. We show, via both quantitative and qualitative metrics, that numerically this method exhibits superior performance to methods based on traditional NMF. It is also more memory efficient and can be tailored to perform online denoising of streaming speech and music. {In the future, one might hope to build upon this research by using more sophisticated loss functions than the one considered here. For instance, one might use the Kullback-Leibler divergence in place of the Frobenius norm or use carefully crafted regularizers to enforce a certain time-frequency structure on the atoms. } 

\bibliographystyle{ieeetr}

\end{document}